\documentstyle[cite,epsf]{article}

\def\And{{\rm and\ }}
\def\Name#1{{\sc #1},}

\begin{document}


\title{Universality and the five-dimensional Ising model}
\author{Henk W.J. Bl\"ote and Erik Luijten\\
        Department of Physics, Delft University of Technology\\
        Lorentzweg 1, 2628 CJ Delft, The Netherlands}
\date{February 17, 1997}

\maketitle

\begin{abstract}
  We solve the long-standing discrepancy between Monte Carlo results and the
  renormalization prediction for the Binder cumulant of the five-dimensional
  Ising model.  Our conclusions are based on accurate Monte Carlo data for
  systems with linear sizes up to $L=22$. A detailed analysis of the
  corrections to scaling allows the extrapolation of these results to
  $L=\infty$.  Our determination of the critical point, $K_c=0.1139150$ (4), is
  more than an order of magnitude more accurate than previous estimates.
\end{abstract}

\section{Introduction}
A much-debated issue in recent years is the question of universality of the
five-dimensional Ising
model~\cite{bnpy,binder5d,rickw,prl,mon,parisi,comm5d,comm5d-r}. This question
focuses on the value of the renormalized coupling constant $g$ at criticality,
which is related to the Binder cumulant $B$~\cite{binder-cum}. For
$d$-dimensional systems with periodic boundary conditions, hypercubic geometry
and $d \geq d_{\rm u}$, where $d_{\rm u}$ denotes the upper critical dimension,
renormalization theory~\cite{brez-zj} predicts that this cumulant has a
universal value.  However, although $d_{\rm u}=4$ for Ising models with
short-range couplings, Monte Carlo simulations~\cite{bnpy,binder5d} for
five-dimensional Ising systems with linear sizes $3 \leq L \leq 7$ suggested a
different value for $B$.  Large-scale simulations for $5 \leq L \leq
17$~\cite{rickw} corroborated the earlier Monte Carlo result.  As this
controversy might indicate a problem with the renormalization analysis, various
efforts were undertaken to gain additional insight in the nature of the
discrepancy.  First, simulations were carried out for a closely-related class
of systems, namely low-dimensional systems with algebraically decaying
interactions~\cite{prl}.  Provided that these long-range interactions decay
sufficiently slowly, they induce classical critical behaviour even in
one-dimensional systems. Thus they effectively lower the upper critical
dimension.  As these systems are described by the same renormalization
equations as high-dimensional short-range models, the same discrepancy in the
Binder cumulant could be observable. The advantage of examining these
long-range systems is their lower dimensionality, which makes it possible to
simulate a much larger range of system sizes. It might seem that this advantage
is undone by the increase of simulation time due to the larger number of
interacting neighbours, which constitutes the very reason for the mean-field
like behaviour.  However, this latter problem was avoided by a cluster
algorithm for long-range interactions in which the simulation time is
independent of the number of interacting spins~\cite{ijmpc}.  The Binder
cumulant was shown to agree accurately with the theoretical prediction for all
examined systems with $d>d_{\rm u}$ (for $d=1,2,3$). Nevertheless, this did not
completely resolve the existing discrepancy, as the relation between models
with long-range interactions and high-dimensional short-range Ising models is
non-exact. Two subsequent studies actually were concerned with the
five-dimensional model itself. Mon~\cite{mon} studied the finite-size behaviour
of the first and third absolute magnetization moments (normalized by the second
moment to render them dimensionless), $\langle |m| \rangle / \langle m^2
\rangle^{1/2}$ and $\langle |m^3| \rangle / \langle m^2 \rangle^{3/2}$, and
found that the Monte Carlo results for these quantities agreed well with the
theoretically expected values.  Furthermore, he showed that the finite-size
corrections for the fourth moment---which is directly related to the Binder
cumulant---are much larger than for the first and third one, which might
explain the previously found disagreement. The only point of discussion
concerning this study was the nature of the dominant finite-size correction,
see refs.~\cite{comm5d,comm5d-r}.  Next, Parisi and Ruiz-Lorenzo~\cite{parisi}
carried out Monte Carlo simulations for the five-dimensional Ising model using
the Wolff cluster algorithm, which implied a considerable improvement compared
to previous studies.  They also introduced a new quantity, namely the Binder
cumulant evaluated at the ``apparent critical temperature'', defined as the
(size-dependent) temperature where the connected susceptibility takes its
maximum. They showed their Monte Carlo results for this quantity, taken at $4
\leq L \leq 16$, to agree well with the mean-field value.  Unfortunately, the
statistical accuracy of the numerical results for the Binder cumulant at the
critical temperature was not sufficient to allow an extrapolation to the $L=
\infty$ limit, so that the original controversy could not be settled yet.

\section{Simulations}
In this paper we present new Monte Carlo results for the five-dimensional Ising
model. We have carried out simulations for hypercubic systems up to linear size
$L=22$, which corresponds to more than $5 \times 10^6$ spins. Periodic
boundaries were employed.  The results have a high statistical accuracy, which
is required to resolve the various finite-size corrections.  The majority of
the results were obtained on a Cray T3E massively parallel computer at Delft
University.  A total amount of 4000 (one-processor) CPU-hours was invested.
One quarter of the total time was spent on the two largest system sizes,
$L=20,22$.  As in ref.~\cite{parisi}, we used the Wolff cluster
algorithm~\cite{wolff} to suppress critical slowing down.  Samples were taken
at intervals containing a number of Wolff steps approximately equal to the
inverse of the average relative cluster size.  Table~\ref{tab:simul} gives the
details for the various system sizes.

\begin{table}
\caption{Details of the Monte Carlo simulations. The table shows both the
  number of Wolff clusters per sample and the total number of samples taken
  for each system size.}
\label{tab:simul}
\begin{center}
\begin{tabular}{rrr}
\hline
\multicolumn{1}{c}{System size}       & 
  \multicolumn{1}{c}{Clusters/sample} &
  \multicolumn{1}{c}{Million Samples} \\ 
\hline
 2              &    5           & 40    \\
 3              &   10           & 36    \\
 4              &   20           & 21    \\
 5              &   30           & 13    \\ 
 6              &   50           & 13    \\
 7              &   70           & 5.3   \\ 
 8              &  100           & 5.8   \\
 9              &  120           & 3.0   \\
10              &  200           & 2.7   \\ 
11              &  200           & 1.6   \\ 
12              &  250           & 1.9   \\ 
13              &  320           & 0.77  \\
14              &  400           & 0.95  \\ 
15              &  500           & 0.51  \\
16              &  600           & 0.64  \\ 
17              &  700           & 0.38  \\
18              &  800           & 0.32  \\
19              &  900           & 0.29  \\
20              & 1000           & 0.26  \\
22              & 1400           & 0.19  \\
\hline
\end{tabular}
\end{center}
\end{table}

\section{Results and discussion}
The main quantity of interest is the universal $L \rightarrow \infty$ limit 
of the amplitude ratio
\begin{equation}
 Q(T,L) \equiv \frac{\langle m_L^2 \rangle ^2}{\langle m_L^4 \rangle} \;,
\end{equation}
which is directly related to the Binder cumulant $B=-3+1/Q$.  In order to
analyze the finite-size data for $Q(T,L)$ we need a description of the
corrections to scaling.  Above the upper critical dimension, the theory of
scaling differs from that below $d_{\rm u}$, because of the presence of a
so-called dangerous irrelevant variable. This leads to a violation of
hyperscaling.  A detailed discussion of the form of the finite-size scaling
functions is given in ref.~\cite{prl}.  The resulting prediction of
renormalization theory is:
\begin{equation}
 Q(T,L) = \tilde{Q}\left(\hat{t} L^{y_{\rm t}^*}, uL^{y_{\rm i}}\right)
          + b_1 L^{d-2y_{\rm h}^*} + \cdots \;,
\end{equation}
where $\tilde{Q}$ is a {\em universal\/} function, $\hat{t} = t + \alpha
L^{y_{\rm i}-y_{\rm t}}$ and $uL^{y_{\rm i}}$ originates from irrelevant
higher-order contributions in the renormalization equations~\cite{prl}.  $t
\equiv (T-T_{\rm c})/T_{\rm c}$ is the reduced temperature.  The asterisks
indicate that the exponents are modified by the dangerous irrelevant
variable.  Following the notation of ref.~\cite{prl} we have $y_{\rm t}^* =
y_{\rm t} - y_{\rm i}/2$ and $y_{\rm h}^* = y_{\rm h} - y_{\rm i}/4$, where in
turn $y_{\rm t}=2$ is the thermal exponent, $y_{\rm h}=(2+d)/2$ the magnetic
exponent and $y_{\rm i}=4-d$ the leading irrelevant exponent. Thus, $y_{\rm
  t}^*=d/2$ and $y_{\rm h}^*=3d/4$.  The one-loop correction $\alpha L^{2-d}$
in $\hat{t}$ is the so-called shift in the critical temperature, which leads to
a finite-size correction proportional to $L^{y_{\rm i}/2} = L^{2-d/2}$ in
$Q(T,L)$. The term $b_1 L^{-d/2}$ arises from the analytic part of the free
energy and the ellipsis stands for higher-order terms.  Upon expansion of the
scaling formula for $Q(T,L)$ near criticality one finds
\begin{equation}
\label{eq:q-fss}
 Q(T,L) = Q + a_1 \hat{t} L^{y_{\rm t}^*} + 
            + a_2 \hat{t} L^{2y_{\rm t}^*} + \cdots
            + b_1 L^{d-2y_{\rm h}^*} + \cdots
            + c_1 L^{y_{\rm i}} + \cdots \;.
\end{equation}

We have fitted eq.~(\ref{eq:q-fss}) to our finite-size data. All data for $L
\geq 5$ were included in the analysis. In addition to the terms in
(\ref{eq:q-fss}) we also used one cross-term in the expansion, viz.\ $\hat{t}
L^{y_{\rm t}^* + y_{\rm i}}$. The exponents of the correction terms, $y_{\rm
  i}$ and $d-2y_{\rm h}^*$, were kept fixed.  The results are shown in
table~\ref{tab:fit}\@. In the first analysis, one observes that both $y_{\rm
  t}^*$ and $Q$ agree with the theoretical predictions, $y_{\rm t}^*=d/2$ and
$Q = 8\pi^2 / [\Gamma(\frac{1}{4})]^4 \approx 0.456947$. Comparing this to
previous studies, we make the following remarks. The best estimate in
ref.~\cite{rickw} is $Q = 0.489~(6)$, more than five standard deviations from
the renormalization prediction. This value deviates approximately four
(combined) standard deviations from our prediction. Furthermore, the quoted
error margin is of the same order as ours.  Since our data have much smaller
statistical errors, this indicates that less correction terms were taken into
account in ref.~\cite{rickw}.  Indeed, the absence of certain finite-size
corrections was suggested in ref.~\cite{prl} as a possible explanation for the
discrepancy. In refs.~\cite{mon,parisi} the finite-size data were directly
compared to the renormalization prediction for $L= \infty$; no actual
extrapolations to infinite system size were made. Hence, our results now
confirm the renormalization prediction for the Binder cumulant of the
five-dimensional Ising model for the first time, in the sense that the accuracy
of our analysis exceeds the level needed to distinguish between the competing
results for $Q$~\cite{rickw,brez-zj}.

Because the thermal exponent agrees with the predicted value, we have repeated
the analysis with $y_{\rm t}^*$ fixed at this value. The resulting estimate for
$Q$ again agrees with the prediction.  However, comparing the uncertainty in
$Q$ with the error margins quoted in ref.~\cite{prl}, where $y_{\rm t}^*$ was
also kept fixed, we see that the results for $Q$ for the systems with
long-range interactions are even more accurate. Given the large amount of
CPU-time spent on the five-dimensional case, this illustrates how well suited
the low-dimensional long-range systems are for the study of universal
properties above the upper critical dimension.  Finally, in order to lower the
uncertainty in $K_{\rm c}$, we have made a third analysis {\em assuming\/} that
$Q$ takes its theoretical value. All three estimates for the critical coupling
agree within one standard deviation.

\begin{table}[htb]
\caption{Results of the least-squares fits of the universal amplitude ratio
  $Q$. The numbers in parentheses denote the errors in the last digit.}
\label{tab:fit}
\begin{center}
\begin{tabular}{clll}
\hline
  Analysis                          &
  \multicolumn{1}{c}{$y_{\rm t}^*$} &
  \multicolumn{1}{c}{$Q$}           &
  \multicolumn{1}{c}{$K_{\rm c}$}   \\
\hline
1 & 2.46~(9)            & 0.456~(6)                        & 0.1139149~(7) \\
2 & 2.50~\mbox{(fixed)} & 0.454~(5)                         & 0.1139147~(6) \\
3 & 2.50~\mbox{(fixed)} & 0.45694658$\ldots$~\mbox{(fixed)} & 0.1139150~(4) \\
\hline 
\end{tabular}
\end{center}
\end{table}

\begin{figure}
\caption{The Binder cumulant at $K=0.1139100$, where most of our data were
  taken, and $K=0.1139150$, our best estimate of $K_{\rm c}$, versus the system
  size~$L$. The points at the latter coupling were calculated from those at the
  former coupling.  Furthermore the function describing the finite-size
  corrections at criticality (dashed curve) and the $L= \infty$ limit of the
  Binder cumulant (solid line) are shown.}
\label{fig:qcor}
\begin{center}
\leavevmode
\epsfxsize=11cm
\epsfbox{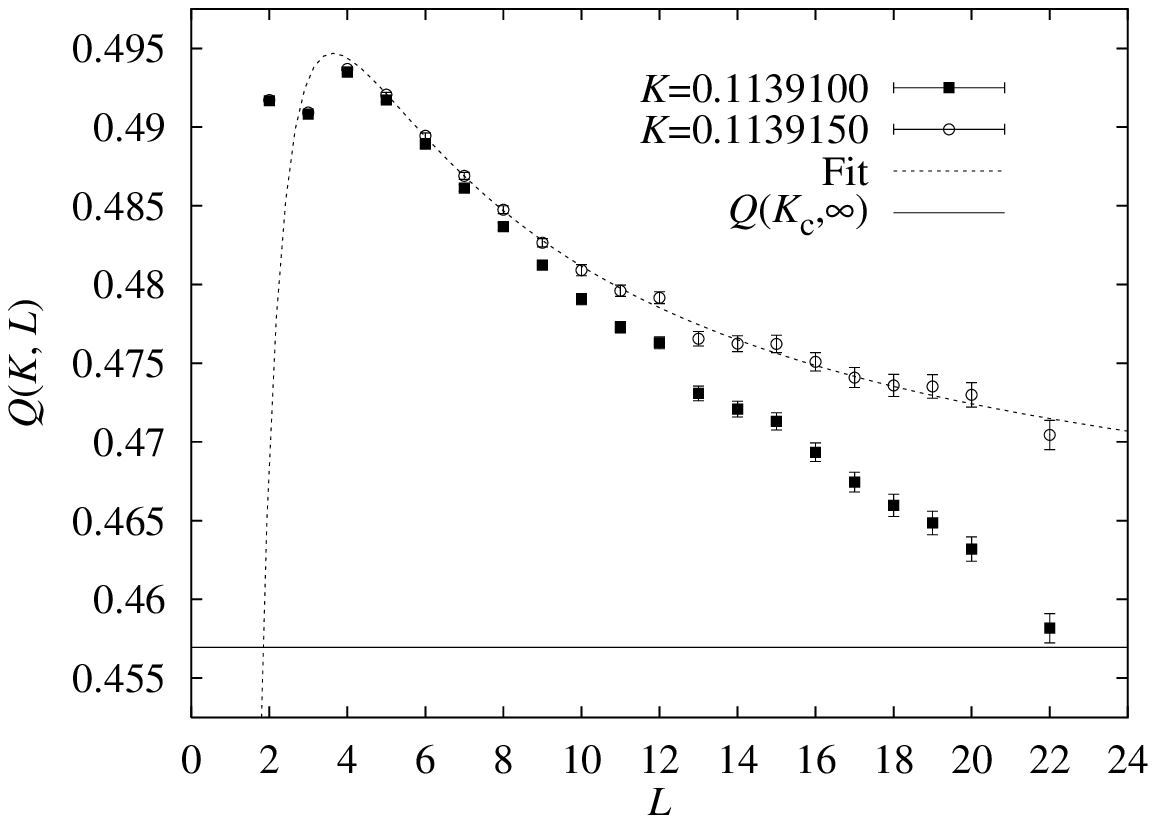}
\end{center}
\end{figure}

In order to gain some insight in the nature of the finite-size corrections
affecting $Q$, we have studied $Q(K_{\rm c},L)$ as function of
$L$.\footnote{For convenience we will from now on use the spin--spin
  coupling~$K$ instead of the temperature~$T$.}  Most of our data were taken
$K=0.1139100$, slightly different from our best estimate for $K_{\rm c}$.
Therefore we have corrected these data for the difference in coupling strength
using eq.~(\ref{eq:q-fss}).  Figure~\ref{fig:qcor} shows both
$Q(K=0.1139100,L)$ and $Q(K_{\rm c},L)$ as function of $L$. This turns out to
be a surprisingly instructive plot. Firstly, one notices that for the larger
values of $L$ the finite-size data for $Q$ are {\em strongly\/} dependent on
the coupling, which is due to the large value of $y_{\rm t}^*$. This implies
that an incorrect estimate of $K_{\rm c}$ has a considerable effect on the
resulting estimate of the Binder cumulant. Secondly, one observes that the
dashed curve indicating the finite-size corrections as predicted by
renormalization theory gives a good description of the data down to system
sizes as small as~4 or~5 (cf.\ refs.~\cite{mon,comm5d,comm5d-r}).  The overall
approach to the $L= \infty$ limit is very slow, given the huge number of spins
in the largest system. Returning to the original discrepancy, we have repeated
the least-squares fits with a smaller number of correction terms.  Apart from
an increase in $\chi^2$, indicating the importance of the higher-order terms,
this leads to a higher estimate of the critical coupling and a correspondingly
higher value for $Q$, although it was by no means as high as the result
in~\cite{rickw}. On the other hand, the shift term $\propto L^{-1/2}$ is not
the dominant term for small $L$, as already suggested in
refs.~\cite{prl,comm5d}, but it is neither negligibly small (in contrast with
the results for systems with long-range interactions).  Naturally, for large
$L$ it will dominate all other corrections.

Unlike the Binder cumulant, the critical coupling was estimated in many
studies. Let us therefore compare our estimate for $K_{\rm c}$ with these
previous estimates (table~\ref{tab:kc}). The early result by Fisher and
Gaunt~\cite{fisher} already has a remarkable accuracy, but the quoted
uncertainty turns out to be almost ten times too small. Other series expansion
results~\cite{guttman,muenkel} agree with our prediction; in particular the
result of Guttmann (which was obtained by fixing the critical exponent $\gamma$
at its mean-field value). Still, the uncertainty in this estimate is more than
an order of magnitude larger than the newest Monte Carlo result. The most
accurate result until now from equilibrium Monte Carlo simulations was obtained
by Parisi and Ruiz-Lorenzo~\cite{parisi} and lies one $\sigma$ below our
estimate. Since this value was obtained with $y_{\rm t}^*$ fixed, the
uncertainty has to be compared to that of the second analysis in
table~\ref{tab:fit}.  Finally, two (coinciding) estimates were obtained by
studing the critical dynamics of the five-dimensional Ising
model~\cite{muenkel,stauffer} for very large system sizes and requiring that
the effective dynamical critical exponent approaches its asymptotic value
$z=2$. These results are also in good agreement with our estimate and the
latter may hence be used to make a more accurate study of the critical dynamics
of the five-dimensional Ising model.

\begin{table}
\caption{Critical couplings for the five-dimensional Ising model as obtained in
  various studies.}
\label{tab:kc}
\begin{center}
\begin{tabular}{rclll}
\hline
Reference       & Year &  $K_{\rm c}$    & Method       & Remarks  \\
\hline
\cite{fisher}   & 1964 &  0.114035 (13)  & series exp.  &  \\
\cite{guttman}  & 1981 &  0.113917 (7)   & series exp.  & $\gamma$ fixed \\
\cite{bnpy},\cite{binder5d} & 1985 & 0.1140 & Monte Carlo & $L \leq 7$ \\
\cite{muenkel}  & 1993 &  0.113935 (15)  & series exp.  &  \\
\cite{muenkel}  & 1993 &  0.11391 (1)    & dynamic MC   & $L \leq 48$ \\
\cite{rickw}    & 1994 &  0.113929 (45)  & Monte Carlo  & $L \leq 17$ \\
\cite{mon}      & 1996 &  0.11389 (13)   & Monte Carlo  & $L \leq 14$ \\
\cite{parisi}   & 1996 &  0.11388 (3)    & Monte Carlo  & $L \leq 16$, 
                                                          $y_{\rm t}^*$ fixed\\
\cite{stauffer} & 1996 &  0.11391 (1)    & dynamic MC   & $L = 112$ \\
This work       & 1997 &  0.1139149 (7)  & Monte Carlo  & $L \leq 22$ \\
This work       & 1997 &  0.1139150 (4)  & Monte Carlo  & $L \leq 22$, $Q$ and
                                                          $y_{\rm t}^*$ fixed\\
\hline
\end{tabular}
\end{center}
\end{table}

\section{Conclusion}
In this paper we have presented numerical results for the five-dimensional
Ising model, in particular for the Binder cumulant and the critical coupling.
Using accurate results for relatively large system sizes we have been able 
to carry out a detailed analysis of the various corrections to scaling. The
results are in full agreement with the predictions of renormalization
theory and hence resolve the long-standing discrepancy for the Binder cumulant.
Furthermore, we reinforced our earlier suggestion that this discrepancy was
caused by the neglect of higher-order finite-size corrections.  A more
elaborate analysis of the critical properties of the five-dimensional Ising
model will be presented elsewhere.

We thank the Centre for High-Performance Applied Computing at Delft University
of Technology for use of the Cray T3E during its testing phase and Prof.\ 
D.~Stauffer for a preprint of ref.~\cite{stauffer}.


\vskip-12pt
\begin{thebibliography}{99}
\itemsep -0.4pt
\bibitem{bnpy}
  \Name{Binder K., Nauenberg M., Privman V. \And Young, A.P.}
  {\it Phys.\ Rev.\ B}, {\bf 31} (1985) 1498.
\bibitem{binder5d}
  \Name{Binder K.} {\it Z. Phys.\ B}, {\bf 61} (1985) 13.
\bibitem{rickw}
  \Name{Rickwardt Ch., Nielaba P. \And Binder K.} 
  {\it Ann.\ Physik (Leipzig)}, {\bf 3} (1994) 483.
\bibitem{prl}
  \Name{Luijten E. \And Bl\"ote H.W.J.} 
  {\it Phys.\ Rev.\ Lett.}, {\bf 76} (1996) 1557; {\bf 76} (1996) 3662.
\bibitem{mon}
  \Name{Mon K.K.} {\it Europhys.\ Lett.}, {\bf 34} (1996) 399.
\bibitem{parisi}
  \Name{Parisi G. \And Ruiz-Lorenzo J.J.}
  {\it Phys.\ Rev.\ B}, {\bf 54} (1996) R3698.
\bibitem{comm5d}
  \Name{Luijten E.} {\it Europhys.\ Lett.}, to be published.
\bibitem{comm5d-r}
  \Name{Mon K.K.} {\it Europhys.\ Lett.}, to be published.
\bibitem{binder-cum}
  \Name{Binder K.} {\it Z. Phys.\ B}, {\bf 43}, 119 (1981). 
\bibitem{brez-zj} 
  \Name{Br\'ezin E. \And Zinn-Justin J.} 
  {\it Nucl.\ Phys.\ B}, {\bf 257} [FS14] (1985) 867.
\bibitem{ijmpc}
  \Name{Luijten E. \And Bl\"ote H.W.J.}
  {\it Int.\ J. Mod.\ Phys.\ C}, {\bf 6} (1995) 359.
\bibitem{wolff}
  \Name{Wolff U.} {\it Phys.\ Rev.\ Lett.}, {\bf 62} (1989) 361.
\bibitem{fisher}
  \Name{Fisher M.E. \And Gaunt D.S.}
  {\it Phys.\ Rev.}, {\bf 133} (1964) A224.
\bibitem{guttman}
  \Name{Guttmann A.J.} {\it J. Phys.\ A}, {\bf 14} (1981) 233.
\bibitem{muenkel}
  \Name{M\"unkel Ch., Heermann D.W., Adler J., Gofman M. \And Stauffer D.}
  {\it Physica A}, {\bf 193} (1993) 540.
\bibitem{stauffer}
  \Name{Stauffer D. \And Knecht R.}
  {\it Int.\ J. Mod.\ Phys.\ C}, {\bf 7} (1996) 893.
\end{thebibliography}
\end{document}